\documentclass{article}
\usepackage{amsfonts}
\usepackage{amssymb}
\usepackage{amsmath}
\usepackage{amsthm}
\usepackage[english]{babel}
\begin{document}

\title{\bf On hyperbolicity violations\\ in cosmological models with vector fields}

\author{{\bf Alexey Golovnev and Aleksandr Klementev}\\
{\small {\it Saint-Petersburg State University, high energy physics department,}}\\
{\small \it Ulyanovskaya ul., d. 1; 198504 Saint-Petersburg, Petrodvoretz; Russia}\\
{\small agolovnev@yandex.ru, \quad sas5292@yandex.ru}}
\date{}

\maketitle

\begin{abstract}

Cosmological models with vector fields received much attention in recent years. Unfortunately, most of them are plagued with severe instabilities or other problems. In particular, it was noted in Ref. \cite{Esposito} that the models with a non-linear function of the Maxwellian kinetic term do always imply violations of hyperbolicity somewhere in the phase space. In this work we make this statement more precise in several respects and show that those violations may not be present around spatially homogeneous configurations of the vector field.

\end{abstract}

\vspace{1cm}

\section{Introduction}

In the recent years there was an interest towards constructing inflationary models with vector \cite{GMV} or even higher spin fields \cite{GerKe}. Unfortunately, the models of vector inflation were shown to contain many serious pathologies and instabilities, such as uncontrolled growth of anisotropies in models with large fields \cite{GMV2}, longitudinal ghosts \cite{Peloso,me}, an extra degree of freedom from temporal components of the vectors \cite{me} which appears due to non-minimal coupling and shows up only at non-linear level if around a homogeneous background. The major reason for the lack of success was that, in order to ensure the slow roll condition, a large tachyonic mass of the vector fields was required which necessarily makes the longitudinal mode a ghost. It led to the common conclusion that vector inflation is unviable due to these instabilities, see however the Ref. \cite{Mindaugas}.

Subsequent developments have shown that a non-decaying vector field in a rapidly expanding Universe can exist if its energy density is subdominant and a special kinetic coupling to some scalar inflaton is invoked \cite{WKS}. Another very promising possibility is represented by the so-called gauge-flation with non-Abelian vector fields \cite{Shahin}. However, for pure Abelian vector theories it has been proven that, at very best, only extremely fine-tuned models can exhibit reasonably stable behaviour \cite{Esposito,me2}. And in passing, apart from the cosmological results, some new properties of non-standard vector field models were investigated. In particular, it was found \cite{Esposito} that for a vector field with an action of the form $$S=-\int d^4 x\left(f(F^2)+V(A^2)\right)$$ where $F_{\mu\nu}\equiv\partial_{\mu}A_{\nu}-\partial_{\nu}A_{\mu}$ is the usual field strength, the well-posedness of Cauchy problem breaks down somewhere in the available phase space of the model. Having in mind that such models might still be used in cosmology, possibly in conjunction with other mechanisms and scalar fields, we revisit this analysis and show that hyperbolicity violations need not be present around homogeneous vector field configurations which are of potential interest for cosmology. At the same time, we also refine the initial argument a bit, describe the concrete patterns of violation, and show that the argument does not give the full picture of possible causality issues since it is insensitive to potential presence of superluminal modes.

In Section 2 we explain the essence of hyperbolicity violation and provide some additions to the treatment in Ref. \cite{Esposito}. In Section 3 we give more details on how precisely the hyperbolicity is violated in a number of paradigmatic situations. In Section 4 we discuss the problem of cosmological vector fields, and give an illustration and some comments on the no-go statement \cite{Esposito} for this class of models. Finally, in Section 5 we conclude.

\section{Hyperbolicity violation}

To state it again, we are working with Lagrangians of the form
\begin{equation}
\label{basic}
{\mathcal L}=-f(F^2)-V(A^2)
\end{equation}
which implies the following equations of motion
\begin{equation}
\label{eom}
\bigtriangledown_{\mu}\left(f^{\prime}(F^2)\cdot F^{\mu\nu}\right)=\frac12 V^{\prime}\cdot A^{\nu}.
\end{equation}

It is easy to find the principle (second order) part of the differential operator in Eq. (\ref{eom}):
\begin{equation}
\label{diff}
{\mathfrak D}^{\nu}_{\mu}A^{\mu}\equiv\left[f^{\prime}\cdot\left(\delta^{\nu}_{\mu}\square-\partial_{\mu}\partial^{\nu}\right)+4f^{\prime\prime}\cdot F^{\alpha\nu}F_{\beta\mu}\partial_{\alpha}\partial^{\beta}\right]A^{\mu}.
\end{equation}
Hyperbolicity of this system of equations can be addressed in terms of the spectral properties of the principle symbol of the differential operator,
\begin{equation}
\label{symbol}
D^{\nu}_{\mu}(p)\equiv (M^{\nu}_{\mu})^{\alpha\beta}
p_{\alpha}p_{\beta}\equiv\left[f^{\prime}\cdot\left(\delta^{\nu}_{\mu}g^{\alpha\beta}-\delta^{\alpha}_{\mu}g^{\nu\beta}\right)+4f^{\prime\prime}\cdot F^{\alpha\nu}{F^{\beta}}_{\mu}\right]p_{\alpha}p_{\beta}.
\end{equation}
First, one needs to diagonalise $(M^{\nu}_{\mu})^{\alpha\beta}$ with respect to $\mu$ and $\nu$ indices, $(M^{\nu}_{\mu})^{\alpha\beta}\rightarrow G_{(\mu=\nu)}^{\alpha\beta}\cdot\delta^{\nu}_{\mu}$. And then each of the $G^{\alpha\beta}$-matrices should have one negative and three positive eigenvalues (in four-dimensional space-time with "mostly plus" signature) which amounts to being the principle symbol of a hyperbolic operator.

Strictly speaking, ${\mathfrak D}^{\nu}_{\mu}$ is not hyperbolic since the equations of motion imply the well-known constraint
\begin{equation}
\label{constraint}
\bigtriangledown_{\mu}\left(V^{\prime}\cdot A^{\mu}\right)=0
\end{equation}
which reduces the number of dynamical components to three. What we really want is that all three $G^{\alpha\beta}$-matrices for those independent components are of hyperbolic nature. And moreover, following the Ref. \cite{Esposito}, we would abuse the terminology in one more respect. Namely, we will require that $g_{\nu\alpha}G^{\mu\alpha}$ is positive definite which means that not only the equations are hyperbolic but also the negative eigenvector  of $G^{\mu\nu}$ is time-like according to the physical metric $g_{\mu\nu}$. This is a perfectly reasonable abuse because, at the end of the day, we need to prescribe a common Cauchy hypersurface for solving the equations of motion of all physical fields.

Note that in Ref. \cite{Esposito}, the condition $\partial_{\mu}A^{\mu}=0$ was imposed which is valid only for the mass-term (quadratic) potential. In this case we see that $$(M^{\nu}_{\mu})^{\alpha\beta}=f^{\prime}\delta^{\nu}_{\mu}g^{\alpha\beta}+4f^{\prime\prime} F^{\alpha\nu}{F^{\beta}}_{\mu}$$ and the symbol acquires the form of $$D(p)=f^{\prime}p^2 I+4f^{\prime\prime}\left| w \right\rangle\left\langle w\right|$$ where $I$ is the unit matrix, $p^2\equiv p_{\mu}p^{\mu}$, and $w^{\mu}\equiv F^{\alpha\mu}p_{\alpha}$. 

Diagonalisation gives three symbols with $G^{\alpha\beta}=f^{\prime}\cdot g^{\alpha\beta}$ which correspond to an obviously hyperbolic operator $f^{\prime}\square$ with a natural requirement that $f^{\prime}>0$. However, the fourth one, $f^{\prime}p^2+4f^{\prime\prime}\left\langle w | w \right\rangle$, is more interesting:
\begin{equation}
\label{effmetric}
G^{\alpha\beta}=f^{\prime}\cdot g^{\alpha\beta}+4f^{\prime\prime}\cdot F^{\alpha\mu}{F^{\beta}}_{\mu}.
\end{equation}
The matrix $g_{\nu\alpha}G^{\mu\alpha}=f^{\prime}\cdot \delta^{\mu}_{\nu}+4f^{\prime\prime}\cdot F^{\alpha\mu}F_{\alpha\nu}$ has been studied in Ref. \cite{Esposito}, and we will not repeat this analysis here. The final result is that this matrix is diagonalisable with two multiplicity two eigenvalues
\begin{equation}
\label{eigen}
\lambda=f^{\prime}+f^{\prime\prime}F^2\pm f^{\prime\prime}\sqrt{(F^2)^2+(F{\tilde F}})^2
\end{equation}
where $\tilde F$ is the dual field strength tensor, and these eigenvalues can be made negative by an appropriate choice of the field variables for any non-linear function $f(F^2)$.

If we want to make such a statement for vector fields with general potentials $V(A^2)$, the condition $\partial_{\mu}A^{\mu}=0$ can no longer be used. Nevertheless, now we proceed to show that the same analysis can be done with no substantial complications. Indeed, the principal symbol (\ref{symbol}) has the following form
$$D(p)=f^{\prime}p^2 I-f^{\prime}\left| v \right\rangle\left\langle v\right|+4f^{\prime\prime}\left| w \right\rangle\left\langle w\right|$$
where $w^{\mu}\equiv F^{\alpha\mu}p_{\alpha}$, $v^{\mu}\equiv g^{\alpha\mu}p_{\alpha}$ and $\left\langle v | w \right\rangle=0$ due to antisymmetry of $F_{\mu\nu}$. We see that, after diagonalisation, two operators still preserve the trivial form of $f^{\prime}\square$ which only requires $f^{\prime}>0$, one operator corresponds again to the effective metric (\ref{effmetric}) with the same conclusion on hyperbolicity properties, and the last one is given by $f^{\prime}\left(p^2-\left\langle v | v \right\rangle\right)=0$. The latter result is perfectly consistent with the fact that there is a constraint (\ref{constraint}) in the model.

One might probably hope that the problems with hyperbolicity could be mitigated after employing the constraint (\ref{constraint}) which was not taken into account at the level of principle (second order) parts of the differential operator. Unfortunately, this is not the case. Indeed, substituting the constraint (\ref{constraint}) into the operator (\ref{diff}), we have for the principle symbol
$$D^{\nu}_{\mu}(p)\equiv (M^{\nu}_{\mu})^{\alpha\beta}
p_{\alpha}p_{\beta}\equiv\left[f^{\prime}\cdot\left(\delta^{\nu}_{\mu}g^{\alpha\beta}+2 g^{\nu\beta}\frac{V^{\prime\prime}}{V^{\prime}}A^{\alpha}A_{\mu}\right)+4f^{\prime\prime}\cdot F^{\alpha\nu}{F^{\beta}}_{\mu}\right]p_{\alpha}p_{\beta}.$$
Now the diagonalisation is not as trivial as before. The symbol has the form of
$$D(p)=f^{\prime}p^2 I+2f^{\prime}\frac{V^{\prime\prime}}{V^{\prime}}\left(A^{\mu}p_{\mu}\right)\left| v \right\rangle\left\langle A\right|+4f^{\prime\prime}\left| w \right\rangle\left\langle w\right|$$
where $w^{\mu}$ and $v^{\mu}$ are the same as above. However, as $\left\langle v | w \right\rangle=0$, we see that the conjugate matrix still has the $\left\langle w\right|$ eigenvector with the eigenvalue given by the effective metric (\ref{effmetric}). And therefore, the hyperbolicity violation is precisely the same as for the mass-term case.

\section{Patterns of violation}

Now we proceed to understanding the concrete patters of hyperbolicity violations in models given by the Lagrangian (\ref{basic}). Let us first consider the case of spatially homogeneous field $A^{\mu}(t)$ in FRW space-time $ds^2=-dt^2+a^2 (t)d{\overrightarrow x}^2$ which is the most interesting case for cosmology. The field strength tensor contains only the electric part, $$F_{0i}\equiv E_i (t)={\dot A}_i (t),$$ and the temporal component of the vector field $A_0$ is necessarily zero \cite{GMV}. One can evaluate the effective inverse metric (\ref{effmetric}), and the result is
$$G^{\mu\nu}\partial_{\mu}\partial_{\nu}=-\left(f^{\prime}-4f^{\prime\prime}\frac{E^2}{a^2}\right)\partial_t^2+\frac{1}{a^2}\left(f^{\prime}\delta_{ij}-4f^{\prime\prime}\frac{E_i E_j}{a^2}\right)\partial_i \partial_j.$$
We see that in models with $f^{\prime}>0$ and $f^{\prime\prime}<0$ hyperbolicity is never violated around the cosmological solutions, even in the stricter sense of the Ref. \cite{Esposito}. Moreover, the propagation velocity is equal to that of light along the electric field $\overrightarrow E$ and subluminal in orthogonal directions. Indeed, consider a spatial rotation which makes $E_i=E\delta^1_i$. After that we get
$$G^{\mu\nu}\partial_{\mu}\partial_{\nu}=-\left(f^{\prime}-4f^{\prime\prime}\frac{E^2}{a^2}\right)\partial_t^2+\frac{1}{a^2}\left(f^{\prime}-4f^{\prime\prime}\frac{E^2}{a^2}\right)\partial_1^2+\frac{1}{a^2} f^{\prime}\partial_2^2+\frac{1}{a^2} f^{\prime}\partial_3^2$$
which does not allow for superluminal modes. On the contrary, if $f^{\prime\prime}>0$ then the propagation velocities along the $x^2$ and $x^3$ axes are superluminal by a factor of $\sqrt{\frac{1}{1-4\frac{f^{\prime\prime}}{f^{\prime}}\frac{E^2}{a^2}}}$. Note that the eigenvalues in the Eq. (\ref{eigen}) are not sensitive to that. But once the fields become large enough and $4f^{\prime\prime}\frac{E^2}{a^2}>f^{\prime}$, the time and the space coordinate along the electric field exchange their roles which implies two negative eigenvalues of the matrix $g_{\nu\alpha}G^{\mu\alpha}$. Of course, the same conclusion can be reached by examining the general formula for the eigenvalues (\ref{eigen}) with $F\tilde F=0$ and $F^2=-2\frac{E^2}{a^2}$.

Let us have a look at a simple example with the kinetic function $f(F^2)=\frac14 F^2 + \epsilon (F^2)^2$. In this case $f^{\prime}=\frac14+2\epsilon F^2=\frac14-4\epsilon \frac{E^2}{a^2}$ and $f^{\prime\prime}=2\epsilon$. We see that for $\epsilon<0$ the model is always well-behaved around the cosmological background, and the propagation velocities are not exceeding the velocity of light. On the other hand, with $\epsilon>0$ the propagation velocities in directions orthogonal to the electric field are superluminal by a factor of $\sqrt{\frac{\frac14 - 4\epsilon\frac{E^2}{a^2}}{\frac14 - 12\epsilon\frac{E^2}{a^2}}}$. And then, the first eigenvalue pathology to be seen at the high electric field energy densities is the mismatch of temporal directions for $\frac{E^2}{a^2}>\frac{1}{48\epsilon}$. And if we raise the field values even farther beyond the causal pathology limit, then at $\frac{E^2}{a^2}>\frac{1}{16\epsilon}$ the whole thing becomes a ghost due to $f^{\prime}<0$.

Next, assume that the field is purely magnetic with the field strength directed along the $x^2$ axis: $F_{31}=B$, or $A_1=\frac{B}{2} x^3$ and $A_3=-\frac{B}{2} x^1$ with some constant $B$. In this case we readily see that
$$G^{\mu\nu}\partial_{\mu}\partial_{\nu}=-f^{\prime}\partial_t^2+\frac{1}{a^2}\left(f^{\prime}+4f^{\prime\prime}\frac{B^2}{a^2}\right)\partial_1^2+\frac{1}{a^2} f^{\prime}\partial_2^2+\frac{1}{a^2} \left(f^{\prime}+4f^{\prime\prime}\frac{B^2}{a^2}\right)\partial_3^2.$$
If $f^{\prime}>0$ and $f^{\prime\prime}>0$ there are no causal pathologies in the sense of Ref. \cite{Esposito}, though the propagation velocities are superluminal in directions orthogonal to $\overrightarrow B$ by a factor of $\sqrt{1+4\frac{f^{\prime\prime}}{f^{\prime}}\frac{B^2}{a^2}}$. If we set $f^{\prime\prime}<0$ then there is still no pathology as long as $\frac{B^2}{a^2}<-\frac{f^{\prime}}{4f^{\prime\prime}}$, and the propagation velocities are less than that of light. At higher magnetic fields, hyperbolicity is violated with three time-like directions in the effective metric, or in four dimensions, one can say that this particular equation became hyperbolic with respect to the spatial coordinate $x^2$.

Finally, one can study configurations with crossed electric and magnetic fields. For the sake of simplicity, we would assume that the fields are given by $F_{01}=E$ and $F_{31}=B$, and the space-time metric is taken to be Minkowskian, in which case the wave operator (\ref{effmetric}) acquires the form
$$G^{\mu\nu}\partial_{\mu}\partial_{\nu}=-\left(f^{\prime}-4f^{\prime\prime}E^2\right)\partial_t^2-8f^{\prime\prime}EB\partial_t \partial_3+\left(f^{\prime}+4f^{\prime\prime}B^2\right)\partial_3^2+
\left(f^{\prime}+4f^{\prime\prime}(B^2-E^2)\right)\partial_1^2+ f^{\prime}\partial_2^2.$$
Whatever is the sign of $f^{\prime\prime}$, one can always have causality problems with the $x^1$ direction which confirms the general statement of hyperbolicity violation from the Ref. \cite{Esposito}.

\section{Cosmological vector fields}

In this Section, we would briefly discuss the possible implications of these results for the problem of cosmological vector fields, in the approximation of a test field in an expanding Universe. Suppose, the potential term of the vector field is subdominant. Or one can even imagine a non-linear $U(1)$ gauge field with ${\mathcal L}=-f(F^2)$; we only keep a potential term in mind in order to avoid discussions of the gauge freedom. Then a spatially homogeneous vector $A^{\mu}$ might be of relevance for cosmology if we manage to make the scalar invariant $\frac{E^2}{a^2}$ slowly rolling. It requires $E\sim a$, or ${\dot E}\approx HE$ where $H\equiv\frac{\dot a}{a}$ is the Hubble constant. Neglecting the mass, we bring the equation of motion (\ref{eom}) into the form
\begin{equation}
\label{cosmological}
\left(1-4\frac{E^2f^{\prime\prime}}{a^2f^{\prime}}\right){\dot E}+H\left(1+4\frac{E^2f^{\prime\prime}}{a^2f^{\prime}}\right)E\approx 0.
\end{equation}
We see that, in order to make ${\dot E}\approx HE$ for a given instant of time, the ratio $\frac{4\frac{E^2f^{\prime\prime}}{a^2f^{\prime}}+1}{4\frac{E^2f^{\prime\prime}}{a^2f^{\prime}}-1}$ must be driven close to unity which is possible only when $4\frac{E^2\left|f^{\prime\prime}\right|}{a^2f^{\prime}}\gg 1$. As we have shown in the previous Section, for $f^{\prime\prime}>0$ there is a causal pathology for such vector field values. And, even at smaller fields, one can also worry about the superluminal modes in this sector. Amazingly, as on many other occasions, the problems of causality and stability prevent us from obtaining a cosmologically relevant solution.

 For $f^{\prime\prime}<0$, the model is much healthier but one has to be cautious for not to drive the vector field to the ghost sector with $f^{\prime}<0$; recall that $F^2=-2\frac{E^2}{a^2}$ increases with decrease of $\frac{E}{a}$. It is tempting to conclude that there might be a suitable model because we are interested in regimes with slowly rolling argument of $f$, so that we can have enough of expansion without bringing $f^{\prime}$ to negative values, and after that there might be a mechanism to ensure a graceful exit instead of the entrance into the ghost mode. However it is not so easy. 

Let us try a vector field with kinetic function $$f(F^2)=1-e^{-\alpha F^2}$$ with some constant $\alpha>0$. In this case we have $f^{\prime}=\alpha e^{-\alpha F^2}>0$ and $f^{\prime\prime}=-\alpha^2 e^{-\alpha F^2}<0$ throughout the whole phase space. And from Eq. (\ref{cosmological}) we obtain
 \begin{equation}
 \label{exponential}
 {\dot E}=\frac{4\alpha\frac{E^2}{a^2}-1}{4\alpha\frac{E^2}{a^2}+1}\cdot HE
 \end{equation}
 which gives ${\dot E}\approx HE$ if $4\alpha\frac{E^2}{a^2}\gg 1$. Moreover, setting $E(t)=a(t)\cdot r(t)$ with $r(t_0)=\frac{E(t_0)}{a(t_0)}$ we get ${\dot r}=-\frac{2Hr}{4\alpha r^2 +1}\approx-\frac{H}{2\alpha r}$. The latter equation can be solved as $r(t)=\sqrt{r^2(t_0)-\frac{1}{\alpha}\log\frac{a(t)}{a(t_0)}}$. The scalar invariant $\frac{E^2}{a^2}$ rolls as $r^2(t)$, and we see that it has only logarithmic dependence on the scale factor. Therefore a parametrically large number of e-folds is possible in the slowly rolling regime.
 
 Unfortunately, this result is misleading. Let us consider the exact solution for this model. It can be found by changing the variable to ${\mathfrak E}=\frac{E}{a}$ in Eq. (\ref{exponential}) and reads $\frac{E^2}{a^2}e^{4\alpha\frac{E^2}{a^2}}=\frac{const}{a^4}$. We see that $\frac{E}{a}$ rolls slow precisely because $f$ is a very steep function. The energy-momentum tensor is dominated by the $\alpha F^{\mu\beta}F_{\nu\beta}e^{-\alpha F^2}$-term which fastly decays, although keeping the relative anisotropy. Therefore, the scalar invariant $\frac{E^2}{a^2}$ is preserved with a high accuracy but the energy density decays exponentially faster. 

Apparently, this example shows us that one has to look for a function $f$ with large $f^{\prime\prime}$ and small $f^{\prime}$ for which no such striking enhancement of rolling of $f$ compared to $\frac{E}{a}$ is present. Effectively, one can take $f^{\prime}\approx\alpha(F_0^2-F^2)$ with $F_0^2<0$ and $F^2$ approaching $F_0^2$ from the side of  $F^2<F_0^2$.  We deduce  from the equation of motion (\ref{cosmological})  that
$$\partial_t F^2=-\frac{4HF^2}{1+2\frac{f^{\prime\prime}}{f^{\prime}}F^2}.$$
And in our case, $F^2$ would roll slowly in vicinity of $F_0^2$ according to $\partial_t F^2\approx 2H(F_0^2-F^2)$. But at the same time we see that the relative change of small quantity $f^{\prime}$ will be as large as $\sim\frac{1}{a^2}$, and therefore for non-decaying energy density, we need to assume that in the regime of interest the energy-momentum tensor is dominated by $fg_{\mu\nu}$-term which, disappointingly, is perfectly isotropic. This is of course not surprising because otherwise it would have been difficult  to achieve the equation of state parameter close to $-1$, even after averaging over several vector fields. Note also, that  before finally reaching $F_0^2$, the function must be rapidly modified by decreasing the absolute value of $f^{\prime\prime}$ in order to avoid transition to $f^{\prime}<0$ mode right after surviving  the expansion.

There are obviously more opportunities to construct the desired equation of state parameter if the potential term in the Lagrangian is taken into account. However it will definitely require fine-tuning to balance the dynamics with interplay of terms of different nature. We agree with the Ref. \cite{Esposito} that, at best, only a very fine-tuned model could be used in order to have non-decaying cosmological vector fields, without explicit time-dependence of parameters or interactions with scalars. However, it is not enough to discuss that in terms of behaviour of $\frac{E}{a}$ (or $\frac{A}{a}$). First, because such regimes are possible. Second, because they do not ensure the slow roll of the true energy density. 

Note also that one can assume an extremely flat function $f(F^2)$ such that $f$ dominates over  $f^{\prime}F^2$ in an exponentially  large range of $F^2$ values. In this case the equation of  state parameter would be close to $-1$, and the energy density would stay almost constant despite the fast roll of the field. This is reminiscent of  solutions with exponentially flat potential terms, although the latter are stable only with some additional requirements \cite{me2}.

\section{Conclusions}

Non-standard vector fields can provide some interesting new ingredients for cosmological model-building. At low energies, their Lagrangians might flow to the standard ones turning the field content into the usual vector fields which are abundant in nature. In opposite limits, they can be important in the energy budget of the (early) Universe, and hopefully might appear capable of producing stable finite degree of anisotropy. 

Many aspects of cosmological vector fields are poorly understood yet, and far not the least reason is that they bring severe stability problems including the issues of hyperbolicity which deserve further investigation. Note that, interestingly, our hyperbolicity conditions in cosmological backgrounds are somewhat similar to corresponding stability conditions in Horndeski vector-tensor models \cite{Lavinia} with the backround geometry quantities playing the role of the background vector configurations.

We have shown that violations of hyperbolicity in vector field models with non-canonical kinetic functions are not necessarily present around the cosmological backgrounds. Interesting solutions with non-trivial dynamics and absence of causal pathologies are possible, though probably not the ones which could naturally allow for stable anisotropy and/or accelerated expansion. However, the non-canonical vector fields can also be used as parts of more complicated scenarios, with additional scalar fields, extra dimensions, time-dependent backgrounds, et cetera. It remains to be seen which effects they might produce in those setups. Therefore, it seems important to have a correct understanding of their analytical properties.

\vspace{0.5cm}

{\bf Acknowledgements.} AG is supported by Russian Foundation for Basic Research Grant No. 12-02-31214, and by the Saint Petersburg State University grant No. 11.38.660.2013. AG is also grateful to the University of Helsinki, and especially to Mindaugas Kar{\v c}iauskas and Kari Enqvist, for hospitality during his visit when the final parts of this project have been developed.


\begin{thebibliography}{9}
\bibitem{Esposito} G. Esposito-Far{\` e}se, C. Pitrou, J.-Ph. Uzan, {\it Physical Review D} {\bf 81} (2010), 063519; {\it arXiv}: 0912.0481.
\bibitem{GMV} A. Golovnev, V. Mukhanov, V. Vanchurin, {\it Journal of Cosmology and Astroparticle Physics}, JCAP06(2008)009; {\it arXiv}: 0802.2068.
    \bibitem{GerKe} C. Germani, A. Kehagias, {\it Journal of Cosmology and Astroparticle Physics}, JCAP03(2009)028; {\it arXiv}: 0902.3667.
    \bibitem{GMV2} A. Golovnev, V. Mukhanov, V. Vanchurin, {\it Journal of Cosmology and Astroparticle Physics}, JCAP11(2008)018; {\it arXiv}: 0810.4304.
    \bibitem{Peloso} B. Himmetoglu, C.R. Contaldi, M. Peloso, {\it Physical Review Letters}, {\bf 102} (2009), 111301; {\it arXiv}: 0809.2779.
    \bibitem{me} A. Golovnev, {\it Physical Review D}, {\bf 81} (2010), 023514; {\it arXiv}: 0910.0173.
    \bibitem{Mindaugas} M. Kar{\v c}iauskas, D.H. Lyth, {\it Journal of Cosmology and Astroparticle Physics}, JCAP11(2010)023; {\it arXiv}: 1007.1426.
    \bibitem{WKS} M. Watanabe, S. Kanno, J. Soda, {\it Physical Review Letters}, {\bf 102} (2009), 191302; {\it arXiv}: 0902.2883.
\bibitem{Shahin} A. Maleknejad, M.M. Sheikh-Jabbari, J. Soda, {\it Physics Reports}, {\bf 528} (2013), 161; {\it arXiv}: 1212.2921.
    \bibitem{me2} A. Golovnev, {\it Classical and Quantum Gravity}, {\bf 28} (2011), 245018; {\it arXiv}: 1109.4838.
\bibitem{Lavinia} J. Beltr{\' a}n Jim{\' e}nez, R. Durrer, L. Heisenberg, M. Thorsrud, {\it Journal of Cosmology and Astroparticle Physics}, JCAP10(2013)064; {\it arXiv}: 1308.1867.
\end{thebibliography}
\end{document}